\documentclass[sigconf]{acmart}

\def\BibTeX{{\rm B\kern-.05em{\sc i\kern-.025em b}\kern-.08emT\kern-.1667em\lower.7ex\hbox{E}\kern-.125emX}}

\usepackage{xspace}
\newcommand{\Name}{\textit{ISOLDE}\xspace}
\newcommand{\Names}{\textit{ISOLDE}'s\xspace}

\usepackage{todonotes}

\usepackage{adjustbox}
\usepackage{graphicx, color}
\usepackage{subcaption}

\usepackage[utf8]{inputenc}

\usepackage{graphicx, color}

\usepackage{multirow}
\usepackage{tikz}
\def\checkmark{\tikz\fill[scale=0.4](0,.35) -- (.25,0) -- (1,.7) -- (.25,.15) -- cycle;} 

\usepackage{array}
\newcolumntype{$}{>{\global\let\currentrowstyle\relax}}
\newcolumntype{^}{>{\currentrowstyle}}

\usepackage{algorithm}
\usepackage[noend]{algpseudocode}

\newcommand\blfootnote[1]{%
  \begingroup
  \renewcommand\thefootnote{}\footnote{#1}%
  \addtocounter{footnote}{-1}%
  \endgroup
}

\renewcommand\footnotetextcopyrightpermission[1]{}
\begin{document}

\title[Interactive Search and Exploration in Online Discussion Forums]{Interactive Search and Exploration in Online Discussion Forums Using Multimodal Embeddings}

\author{Iva Gornishka}
\affiliation{%
  \institution{Informatics Institute \\ University of Amsterdam}
  \city{Amsterdam} 
  \state{The Netherlands} 
}
\email{iva.gornishka@gmail.com}

\author{Stevan Rudinac}
\affiliation{%
  \institution{Informatics Institute \\ University of Amsterdam}
  \city{Amsterdam} 
  \state{The Netherlands} 
}
\email{s.rudinac@uva.nl}

\author{Marcel Worring}
\affiliation{%
  \institution{Informatics Institute \\ University of Amsterdam}
  \city{Amsterdam} 
  \state{The Netherlands} 
}
\email{m.worring@uva.nl}

\begin{abstract}
In this paper we present a novel interactive multimodal learning system,
which facilitates search and exploration in large networks of social multimedia users. 
It allows the analyst to identify and select users of interest, and to find similar users in an interactive learning setting. 
Our approach is based on novel multimodal representations of users, words and concepts, which we simultaneously learn by deploying a general-purpose neural embedding model. We show these representations to be useful not only for categorizing users, but also for automatically generating user and community profiles. 
Inspired by traditional summarization approaches, we create the profiles by selecting diverse and representative content from all available modalities, i.e. the text, image and user modality. 
The usefulness of the approach is evaluated using artificial actors, which simulate user behavior in a relevance feedback scenario. Multiple experiments were conducted in order to evaluate the quality of our multimodal representations, to compare different embedding strategies, and to determine the importance of different modalities. 
We demonstrate the capabilities of the proposed approach on two different multimedia collections originating from the violent online extremism forum Stormfront and the microblogging platform Twitter, which are particularly interesting due to the high semantic level of the discussions they feature.

\end{abstract}



\maketitle

\blfootnote{The paper is based on the thesis of Iva Gornishka, titled "Interactive Search and Exploration in Social Multimedia Networks", University of Amsterdam, 2018.}

\section{Introduction} \label{sec:intro}

In recent years much of our communication is happening on social multimedia platforms, which allow users to connect with others and to exchange ideas and content about topics of their interest. Such fora commonly host lengthy discussions and fierce debates about social issues, and they have become a widely adopted place for cooperation, activism, promotion of different ideologies, and organization of offline events and activities all over the world. Furthermore, they have become invaluable for members of various social movements and groups of like-minded people, since they do not only bring people with shared views and interests together, but also create a feeling of belonging to a community. Still, as pointed out by \citet{conway2016determining}, even basic descriptive or explanatory research is often missing when it comes to e.g. determining the role of Internet in promoting the ideologies and processes with a potentially cataclysmic societal impact, such as violent extremism and radicalization. One of the main factors hindering domain experts from social sciences fields and law enforcement agencies is a lack of multimedia analytics tools facilitating insight gain into the role of users and groups in dynamic online communities. Tools should be developed that not only model users' interactions with the platform and each other, but go beyond to incorporating analysis of the heterogeneous digital content they consume.

\begin{figure}[t!]
\centering
  \includegraphics[width=\columnwidth]{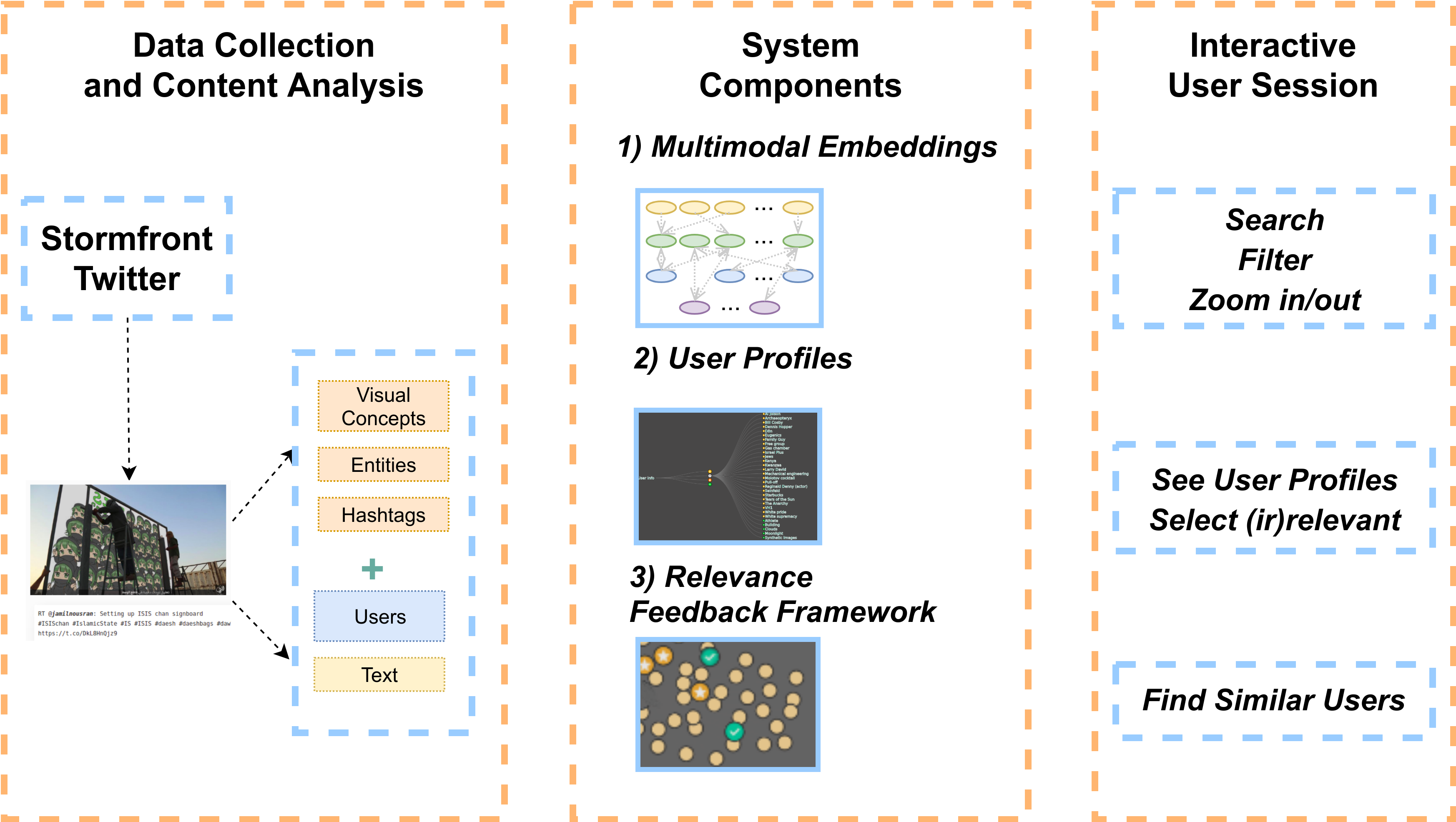}
  \caption[\Name: High-level system overview]{High-level overview of \Names full preprocessing pipeline, system components and interactive user sessions}
  \label{fig:system-td}
\end{figure}

In this paper we design \Name, a novel analytics system which aids domain experts in analyzing large collections of social multimedia users. Its main purpose is to enable easy, on the fly categorization of the users into analytic categories and the discovery of new users of interest, in order to assist annotating large datasets for frequently changing communities. In addition, the proposed system needs to facilitate search and exploration in the large and heterogeneous multimedia collections that online discussion forums create. Lastly, the system should fulfill basic requirements for multimedia analytics systems --- as outlined by \citet{zahalka2018blackthorn}, some of the most important being \textit{interactivity}, \textit{scalability}, \textit{relevance} and \textit{comprehensibility}. 

Thus, we identify several components essential for the system: 
\begin{itemize}
\item First, compact but meaningful multimodal content representations are needed to ensure the interactivity of the system and the relevance of the produced results. These representations need to be suited for diverse and heterogeneous collections, incorporating available information from the text, visual content and member user interactions. 
\item Second, multimodal user and community summaries (profiles) should allow for easy navigation through the space and support insightful analysis of the different social networks and their members. 
\item Finally, the core component of the system should be an interactive learning framework, allowing the domain experts to tailor categorization to their needs. 
\end{itemize}
Based on these technical prerequisites and the common information needs of the domain experts, we develop a prototype of such a system, which contains the following main novel contributions:
\begin{itemize}
\item Compact multimodal user and content representations, which can serve as the basis for visualization, interactive learning, categorization and profiling.
\item Investigation into the effect these embeddings have on different aspects of system performance. Namely, while the effects of text embeddings, such as word2vec, glove and fasttext, on standard evaluation metrics is relatively well studied, this is not the case for multimodal embeddings. 
\item Analysis of a number of use cases to identify in which situations different modalities, modality fusion approaches, traditional vector-space models and multimodal embeddings yield optimal performance.
\end{itemize}

\section{Related Work} \label{sec:relwork}

Over the last decades, analyzing multimodal data and its underlying structure has
become a subject of many studies, and a number of excellent multimedia representation approaches, visualization tools and analytics systems have been proposed to aid domain experts in their research. While some of them were specifically developed to support network analysis tasks, others, although not directly applicable, are characterized by plenty of useful features and could serve as a good starting point for developing a novel analytics system for search and exploration in large collections of social multimedia users. In this section we analyze those likely to satisfy requirements identified in the Introduction.

\subsection{Social Multimedia Representation} \label{sec:smr}

Deriving a joint representation from different modalities associated with a multimedia item has been a long-standing research question in cross-media retrieval \cite{Wang:2017:ACR:3123266.3123326,Wu:2018:LSS:3240508.3240521,Qi:2017:OCS:3123266.3123311,Yang:2016:ZHV:2964284.2964319,Chen:2018:DUC:3240508.3240627}. The main idea behind such approaches is learning a common space to which different modalities, usually text and visual, can be mapped and directly compared. They yield excellent performance in retrieval across different modalities. We conjecture, however, that such approaches are not directly applicable for modeling the complex relations within social multimedia networks. It requires going beyond text and visual content, which are usually well-aligned, to include user interactions with the online platform and each other.

Inspired by the success of word embedding techniques, such as word2vec \cite{NIPS2013_5021}, glove \cite{D14-1162} and fasttext \cite{fasttext:2017}, graph and general purpose embeddings for multimodal data started to emerge \cite{Grover:2016:NSF:2939672.2939754,DBLP:conf/aaai/WuFCABW18}. These approaches were proven effective in a variety of tasks, such as query-by-example retrieval and recommendation. Although they have not been designed or tested on visual content, due to their demonstrated potential for modeling heterogeneous data collections, we consider them a solid base for building multimodal user representations. 

\subsection{Graph Visualization Tools} \label{sec:gv}

Through the years many researchers have addressed the problem of bringing graph visualization to non-technical users with simple tools such as NodeXL \cite{smith2009analyzing}. 
More elaborate systems such as GraphViz \cite{ellson2004graphviz} make it possible to visualize large graphs with multiple different layouts, and even provide simple methods for reducing the visual clutter caused by the huge amount of vertices and edges. 
The main disadvantage of such drawing packages is that they are usually static and do not provide any analysis of the collection. 

More advanced visualization tools make use of various algorithms in order to improve the readability of the graphs, to provide better overview and to enable insight gain. PIWI \cite{yang2013piwi} integrates multiple community detection algorithms to present the user with tag words describing each community. Newdle \cite{yang2010newdle} applies graph clustering with a similar purpose. It allows the user to adjust the algorithm parameters, and provides search functionality based on the cluster tags, as well as simple tag analysis.

One of the most commonly used visualization tools to date is Gephi - an open source software for graph and network analysis on complex datasets \cite{bastian2009gephi}. Gephi does not only allow for interactive exploration of the collection, but also provides filtering, manipulating and clustering functionality, which makes it a great tool for dynamic network visualization. 

While all of the aforementioned tools are great for discovering the overall structure of the collection, they only provide very basic analysis of the content and the individual items, which is mainly done in a pre-processing step. Furthermore, search capabilities are limited and the systems' lack of flexibility does not easily enable more complicated analytic tasks. Last but not least, none of these tools have been developed with the problems of heterogeneous multimedia collections in mind.

\subsection{Multimedia Analytics}

Aggregated search engines such as CoMeRDA \cite{comerda} provide excellent search and filtering functionality in large multimedia databases, but fail to give an overview of the whole collection and the underlying relationships between items. Furthermore, limiting the user to a set of filters and the need to formulate a query make identifying relevant content and serendipitous discovery particularly difficult when the user does not have a well-defined information need.

Multimedia analytics systems, such as Multimedia Pivot Tables \cite{7579240}, ICLIC \cite{van2016iclic}, and Blackthorn \cite{zahalka2018blackthorn} facilitate search and exploration in large collections of multimedia data as well as interactive multimodal learning. Blackthorn, for example, compresses semantic information from the visual and text domain and learns user preferences on the fly from the interactions with the system in a relevance feedback framework. Serving as the epicenter of research on interactive multimedia retrieval, the initiatives such as Video Browser Showdown produced a number of excellent analytics systems \cite{8352047}. For example, vitrivr system owes it good performance in interactive multimedia retrieval to an indexing structure for efficient kNN search~\cite{10.1007/978-3-030-05716-9_55}. Similarly, SIRET tool facilitates interactive video retrieval using several querying strategies, i.e. query by keywords, query by color-sketch and query by example image \cite{10.1007/978-3-319-73600-6_44}. Following a different approach, Vibro system relies on a hierarchical graph structure for interactive exploration of large video collections \cite{10.1007/978-3-319-27674-8_43}. Yet, none of these systems were designed with the analysis of social graphs in mind. 

Finally, while PICTuReVis \cite{doi:10.1111/cgf.13188} facilitates interactive learning for revealing relations between users based on their patterns of multimedia consumption, it is not designed for search and exploration of large social multimedia networks, but rather forensic analysis of artifacts from e.g. confiscated electronic devices, featuring a limited number of users.

\subsection{Summary of System Requirements} \label{sec:sor} 

In Introduction we identified several system components essential for our use case. Considering the strong and weak points of existing systems, here we make an inventory of additional useful features an ideal system for analysis of large social multimedia networks should posses. 

\begin{description}
\item[Overview]: Allow user to get a grip of the whole collection 
\item[Network Analysis] -- facilitate finding groups and communities of interest
\item[Content Analysis] -- provide deeper analysis of the content of the items in the collection
\item[Online Learning] -- allow on the fly categorization of items in the collection 
\end{description}

Table~\ref{table:related-work} summarizes how different systems discussed in this section fare with regard to these requirements.  

\newcommand{\yep}{\checkmark}
\newcommand{\mnah}{$\thicksim$}

\begin{table}[!t]
\centering
\caption{Existing visualization tools and analytics systems.}
\resizebox{\columnwidth}{!}{%
\begin{tabular}{|l|c|c|c|c|}
\hline
 \multicolumn{1}{|m{2.1cm}|}{\centering \textbf{System/Tool}}
& \multicolumn{1}{m{1.3cm}|}{\centering \textbf{Overview}}
& \multicolumn{1}{m{1.2cm}|}{\centering \textbf{Network \\ Analysis}}
& \multicolumn{1}{m{1.2cm}|}{\centering \textbf{Content \\ Analysis}}
& \multicolumn{1}{m{1.2cm}|}{\centering \textbf{Online \\ Learning}}\\ \hline
NodeXL \cite{smith2009analyzing}	&
\yep 	&	
 	&	
	&	 
\\ \hline

GraphViz \cite{ellson2004graphviz}	&
\yep 	&	
 	&	
	& 
\\ \hline

PIWI \cite{yang2013piwi}	&
\yep 	&	
\yep 	&	
 	& 
\\ \hline

Newdle \cite{yang2010newdle} 	&
\yep 	&	
\yep 	&	
	& 
\\ \hline

Gephi \cite{bastian2009gephi}	&
\yep 	&	
\yep	&	
 	& 
\\ \hline

CoMeRDA \cite{comerda} &
	&	
	&	
\yep & 
\\ \hline

Blackthorn \cite{zahalka2018blackthorn}	&
	&	
	&	
\yep	& 
\yep   	
\\ \hline 

vitrivr \cite{10.1007/978-3-030-05716-9_55}	&
\yep	&	
	&	
\yep	& 
\\ \hline

SIRET \cite{10.1007/978-3-319-73600-6_44}	&
\yep	&	
	&	
\yep	& 
\\ \hline

Vibro \cite{10.1007/978-3-319-27674-8_43}	&
\yep	&	
	&	
\yep	& 
\\ \hline

PICTuReVis \cite{doi:10.1111/cgf.13188}	&
\yep	&	
	&	
\yep	& 
\yep   	
\\ \hline

\hline
\Name	&
\yep	&	
\yep	&	
\yep	& 
\yep   	
\\ \hline
\end{tabular}
}
\label{table:related-work}
\end{table}

\section{Approach Overview} \label{sec:system}
In this work, we propose \Name{} -- a novel multimodal analytics system which allows domain experts to explore large collections of social multimedia users, to identify users of interest and find more similar ones in an interactive framework. Our goal goes beyond modeling social networks shaped by user interactions with the online platform and each other, to discovering ``topical communities'' based on similarities in multimedia content the users consume.   
In this section we provide a high-level overview of \Name, its main components and their importance for the user's interactive sessions.

\Names full pipeline is conceptually depicted in Figure \ref{fig:system-td} and it consists of three main phases: data preprocessing, preparation of the individual system components and finally -- the main purpose of the system -- the interactive user sessions. 

\subsection{Data preprocessing}

The first step is collecting social multimedia data and analyzing the content in order to provide additional context from the text, visual, or any other available modality. 
from all images, as well as entities from the text. 
Afterwards, the annotated data is indexed in order to provide search and filtering functionality. 
Section~\ref{sec:data} discusses in details our approach to data collection and analysis, however, \Names framework is general and can easily be adapted to any social multimedia platform or content analysis approach. 

\subsection{System Components}

The following phase in \Names pipeline is setting up all individual components of the system. 
First, we need user representations which will be later used to visualize the users and to classify them. While any standard technique can be used to represent the users, our proposed approach is based on a novel general-purpose neural embedding model which allows us to learn representations not only for users, but also for the multimodal content \cite{wu2017starspace}. 

Next, user and community summaries (profiles) are required to reduce the amount of content that needs to be processed by the interacting user and to support insightful analysis of the different social networks and their members. Our approach to creating such summaries is inspired by standard summarization techniques and relies on identifying diverse and representative items from all modalities. This process is enabled by our multimodal content representations.

Lastly, an interactive learning component enables the discovery of new users of interest based on previously provided examples. 

\subsection{Interactive user sessions}

Finally, all components of the system are combined into a single interactive user interface. From the point of view of the interacting user (i.e. actor), analytic sessions consist of the following steps:

\paragraph{Step 1: Collection overview, search and exploration.}
The actor is initially presented with an aggregated overview of all users in the collection, grouped by their topical similarities. 
They can start interacting with the collection, easily navigating by zooming in and out, filtering the items and searching by text queries.
Automatically created sub-community profiles enable quick understanding of the space and the overall structure of the collection.

\begin{figure}[t!]
\begin{center}
  \includegraphics[width=\columnwidth,keepaspectratio]{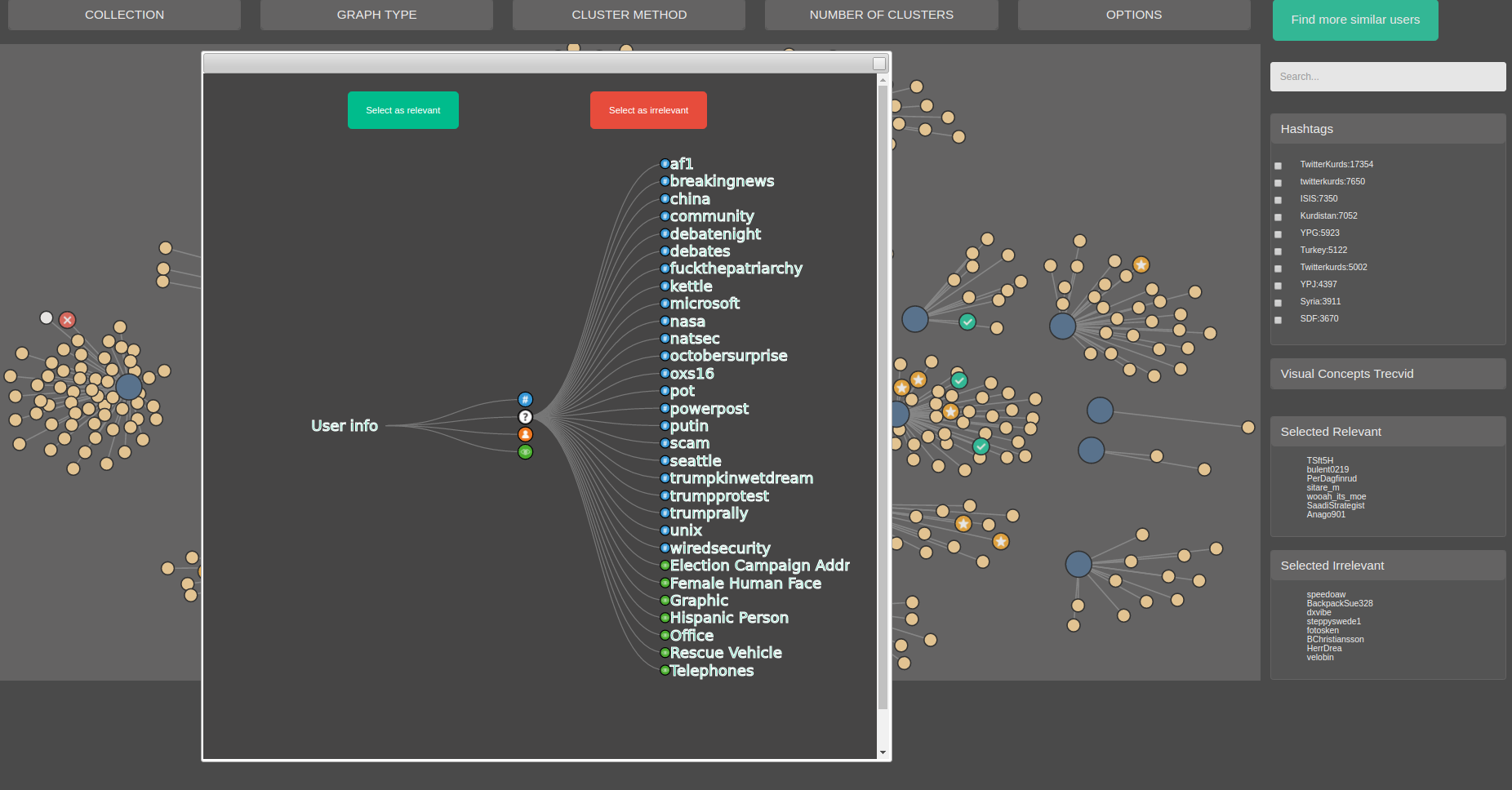}
  \caption[\Name: Interactive Session View]{A screenshot of \Names interface with the detected topical communities in the background and automatically generated user profile in the foreground.}
  \label{fig:system-screenshot-overview-popup}
\end{center}
\end{figure}

\paragraph{Step 2: User interactions and profiling.}
Once potential users of interest are identified, the actor can inspect their automatically generated multimodal user profiles, and mark them as relevant or irrelevant (Figure \ref{fig:system-screenshot-overview-popup}). 

\paragraph{Step 3: Identifying similar users.}
When the actor has marked a number of users as relevant, she can look for more similar users. At this moment, an interactive classifier is trained in real time and used to score each item (i.e. users) in the collections. 
All user nodes are colored in order to reflect the scores, and the top-N users considered most relevant by the system are highlighted in the interface.
Although it is preferable that the actor explicitly indicates the relevance of the top-N highlighted results, they are not obliged to do so and can simply continue with another action.

~\\
Throughout the whole session, the interacting user can seamlessly switch between the aforementioned actions and continue iterative refinement of their preferences until satisfied with the outcome.

\section{Data Collection and Analysis} \label{sec:data}

\subsection{Datasets}

We demonstrate the potential of the proposed system using two different datasets, related to the violent online extremism domain and originally collected for the EU VOX-Pol Project\footnote{\url{http://www.voxpol.eu/}}.

\subsubsection*{Stormfront}
The first collection consists of around 2 million posts from the white nationalist, white supremacist, antisemitic neo-Nazi Internet forum Stormfront. 
After disregarding all posts of suspended users and users with less than 3 posts, the final dataset contains posts from 40 high-level categories, generated by $29.279$ users in the period 2001-2015.
For the purpose of this research, we distinguish between two different subsets of categories -- the \textit{General chapters} (such as \textit{Politics \& Continuing Crises} and \textit{Dating Advice}), containing discussions on various topics, and the \textit{National Chapters} (such as \textit{Stormfront Italia} and \textit{Stormfront Britain}), which are commonly frequented by users interested in the specific geographic area and topics related to it.

\subsubsection*{Twitter}
The second collection consists of almost 2.7 million messages from the microblogging platform Twitter\footnote{\url{http://www.twitter.com/}}. The tweets are in 54 languages and were shared by $9.501$ users involved in discussion about different extremist ideologies. The collection was crawled in 4 separate cycles using Twitter's Search and Streaming APIs\footnote{\url{https://developer.twitter.com/en/docs.html}}, and contains the following subsets of tweets: (1)~\textit{XFR} -- the timelines of 174 users confirmed by social scientists to be supporters of far right ideologies; (2)~\textit{Jihadi} -- the content generated by 86 users confirmed by social scientists to be jihadis (ISIS supporters in particular); (3)~\textit{TwitterKurds} -- tweets collected by using the seed hashtag \#TwitterKurds, as well as variations of the hashtag; (4)~\textit{TwitterTrolls} --  tweets collected by using a number of seed hashtags provided by the social scientists interested in the analysis of \textit{trolling behavior}, an emerging strategy in countering violent online political extremism. In this study we omit all tweets of people who had less than 3 posts in the original dataset.

\subsection{Content Analysis}

When analysing social multimedia data, scientists usually start by annotating the content -- they commonly label the individual posts and images with concepts that describe them. This process is often done manually, which makes it labor intensive and time consuming. In order to mimic this process, we automatically extract concepts from the textual and visual domain as follows.

\subsubsection*{Entity linking}

In the preprocessing step, we extract entities (i.e. topics, people, organizations and locations) mentioned in the Stormfront posts using the Semanticizer \cite{semanticizer:2012, graus2014semanticizing}, which links text to English Wikipedia articles. The process resulted in $65.240$ unique entities, which are usually much easier to interpret than alternatives such as latent topics. 

Although the Semanticizer is based on an approach for adding semantics to microblog posts, and tweets in specific \cite{meij2012adding}, we argue that semantic linking might result in sparse and noisy annotations. 
Stormfront's posts are predominantly well-formed and grammatically correct, as opposed to Twitter's content which contains relatively short messages, use of slang and non-standard abbreviations. Yet, the Semanticizer already fails to detect simple entities in such clean text. 
Furthermore, our Twitter dataset contains around 100K posts which only consist of an image, url, hashtags, tags of other people or a combination of those, hence, no other text at all is available for analysis.
Thus, for the Twitter collection the choice was made to use hashtags in order to provide additional context for the text modality - hashtags, just like entities, commonly refer to named entities, and are commonly used by users to self-label their content, as noted by \citet{teevan2011twittersearch}.

\subsubsection*{Visual concepts}
For each image in both datasets we extract $346$ TRECVID semantic 
concepts\footnote{Full list of the 346 TRECVID concepts (selected from the original set of 500 TRECVID concepts for the TRECVID 2011 Semantic indexing task) is available at \url{https://www-nlpir.nist.gov/projects/tv2011/tv11.sin.346.concepts.simple.txt} \\ More information about the task itself is available at \url{https://www-nlpir.nist.gov/projects/tv2011/}}
as described by \citet{SnoekPTRECVID2013}.
A large number of TRECVID concepts are related to intelligence and security applications, and have been shown more useful for categorizing extremism content than the semantic concept detectors trained on general-purpose image collections such as ImageNet \cite{rudinac2017multimodal}.
After obtaining confidence scores per image for all of the concepts, we only select the top $5$ highest-ranking concepts to represent it. 

~\\
Finally, it is important to mention that one could apply different approaches to providing additional information from the text and visual modalities. Our previous work included annotating both Twitter and Stormfront posts and users with extracted LDA topics \cite{blei2003latent}, but as already mentioned, they are harder to interpret than entities. We also made use of Indri \cite{strohman2005indri} in order to annotate Stormfront users with a predefined list of domain-specific topics, however, for this research, we chose to make use of more general approaches applicable to any social multimedia platform in order to showcase the vast applicability of our system and easily compare performance on our two datasets.

\section{User Representation}

\subsection{Unimodal representation and early fusion}

As a baseline approach to representing a user based on all modalities (text, visual concepts, entities and hashtags) we use TFIDF representations \cite{salton1988term}.
Due to the specifics of our automatic approaches to extracting visual concepts and entities from the posts, some concepts and entities are extracted particularly often -- this could present an issue for alternatives such as bag of concepts representation, but is naturally handled by TFIDF representations.

In order to represent a user in the textual modality, we treat all the posts generated by the user as a single document $d$. Then, for each term $t$ in the vocabulary:

\begin{equation}
tfidf(t, d) 
 = tf(t, d) \times ( \log \frac{1 + n_d} {1 + d(t)} + 1) \label{eq:tfidf}
\end{equation}
where $tf(t, d)$ is the term frequency of $t$ in document $d$, $n_d$ is the total number of documents and $d(t)$ is the number of documents containing $t$. 
In a similar fashion, using the standard definition of TFIDF, we separately create user representations in all other modalities by simply treating all visual concepts, entities and hashtags as terms -- that is, $tf(t, d)$ in equation \ref{eq:tfidf} is the number of times a concept $t$ was used by the user and $d(t)$ is the number of users that used the same concept $t$.

In this manner, we produce user representations having the dimensionality of the corresponding vocabulary -- that is number of words or distinct concepts.
Such high-dimensionality vectors are not suitable for interactive systems like \Name since they increase the computational time needed for classification, and hence the amount of time that the user waits for a response from the system. 
Therefore, we reduce the dimensionality of the obtained unimodal user representations using PCA \cite{martin1979multivariate}. 

Preliminary experiments showed that in our evaluation setup early fusion clearly outperforms late fusion approaches independent of the deployed fusion technique. 
Thus, in all future experiments with TFIDF representations based on multiple modalities, we assume the representations have been normalized and concatenated in advanced. 

~\\
However, separately modeling the individual modalities does not only yield sparse representations in some of them, but it also fails to capture important dependencies between them. When analyzing social multimedia data, analysts are often interested in the co-occurrence of specific topics and the relations between the different modalities within the same posts -- phenomena which are not capture by simple unimodal representations which essentially aggregate all of the content produced by the user. 

\subsection{Multimodal neural embeddings} \label{sec:neural-embeddings}

In order to overcome some of the problems posed by unimodal user representations, we propose learning multimodal user and content representations using StarSpace \cite{DBLP:conf/aaai/WuFCABW18} - a general-purpose neural embedding model which was recently shown to be effective for a variety of tasks. While most of these tasks can be adapted in order to learn user representations, we chose to deploy a multilabel text classification task,
which allows us to simultaneously learn embeddings not only for users, but also for words, entities, hashtags and visual concepts. We expect such embeddings to be general enough for solving a wide range of tasks.

Formally, StarSpace receives as input document-label pairs $(a, b)$ and minimizes the loss

\begin{equation*}
\sum_{\substack{(a, b) \in E^+ \\ b^- \in E^-}} 
L^{batch}(sim(a, b), sim(a, b^-_1), \cdots, sim(a, b^-_k))
\end{equation*}

where $sim$ is cosine similarity; $L^{batch}$ 
\textit{}is hinge loss and the negative labels $b^-_i$ are generated using $k$-negative sampling strategy; documents are represented as bag-of-words and their embedding is simply the sum of the word embeddings in it.

\newcommand{\figEmbeddingTraining}{
\begin{figure}[t!]
\begin{center}
  \includegraphics[width=1\columnwidth,keepaspectratio]{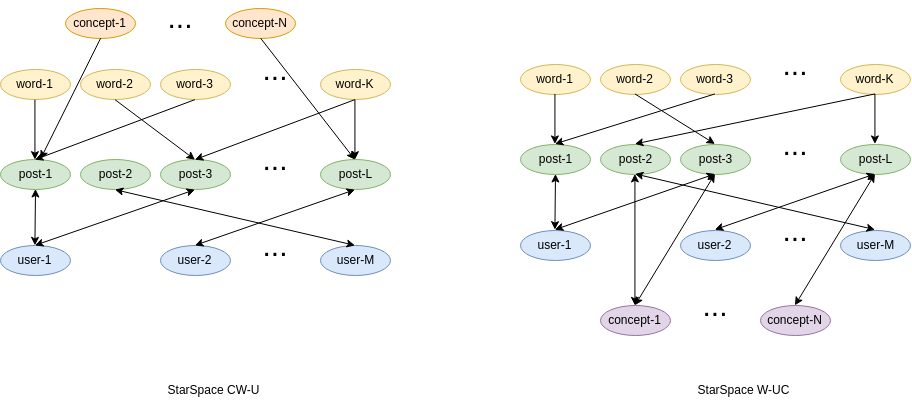}
  \caption[Setups for creating StarSpace embeddings]{Different setups for creating user and content representation using a general purpose neural embedding.}
  \label{fig:embeddings-training}
\end{center}
\end{figure}
}

\figEmbeddingTraining

~\\
While there are multiple possible approaches to training the user and content embeddings through a multilabel text classification task, in the scope of this research we have chosen to compare two different setups, conceptually depicted on Figure \ref{fig:embeddings-training}. In both setups we generate training examples per post and assign the corresponding user as a positive label. However, the setups differ in two ways - first, in the way we generate the input documents for every example (post), and second, in whether or not we add additional labels next to the user label.

StarSpace \textbf{CW-U} trains the model with two separate examples per post - one containing the bag of concepts associated with the post (\textbf{C}), and one containing the bag of words (\textbf{W}). For both examples, the user (\textbf{U}) is a single assigned label.

In StarSpace \textbf{W-UC} setup, examples consist again of the bag of words (\textbf{W}), however every post is labeled not only with the user (\textbf{U}), but also with each of the associated concepts (\textbf{C}). In this way, we put more importance on the concepts -- such a setup implicitly minimizes the distance between users and the concepts that they commonly use, by simultaneously minimizing the distance between user and a post, and each of the concepts and the post. 

\subsection{Profiles}

One of the main challenges with analyzing social media data is the amount of content that needs to be processed. As recently noted by \citet{rudinac2018rethinking}, efficient summarization approaches are needed to facilitate exploration in such large and heterogeneous collections. 

Given the diversity of available social media platforms, we identify the following requirements for a desired summarization approach: first, it needs to be able to summarize content from multiple (possibly unrelated) documents (posts); the quality of the summaries should not depend on the length or content of the individual documents; the summarization process needs to simultaneously handle all available modalities so that it can capture important dependencies between them; and last but not least, it needs to be able to summarize content even in the absence of one or many of the possible modalities.

Our multimodal embeddings make it possible to directly compare items across the different modalities, which allows us to simultaneously pick representative and diverse content from all of them. We generate user profiles by following Algorithm \ref{alg:profiles}.
We start by collecting all possible items that can be included in the summary -- that is the union of $W_U$ (the set of words used by the user), $C_U$ (visual concepts, entities or hashtags extracted from their posts) and $R_U$ (the set of users that $U$ retweeted (for Twitter) or replied to (for Stormfront)).

Next, in an iterative manner we pick items from the set to be included in the summary until we reach a desired number of items. At every iteration, we start by assigning each item $i$ multiple scores which reflect the following criteria:

\begin{itemize}
\item \textbf{User Representativeness:} $SU_i = \#\text{times it was used by } i$ --- pre\-ferable items are commonly used by the user

\item \textbf{Semantic Representativeness:} $SR_i = \sum_{j \in S} dist(i, j)$ --- pre\-ferable items have low total distance to all other items

\item \textbf{Diversity:} $SD_i = \sum_{j \in P} dist(i, j)$ --- pre\-ferable items have high total distance to previously selected items 

\end{itemize}

We use the scores in order to produce $3$ different rankings of the items -- based on ascending $SR$ scores and descending $SD$ and $SU$ scores, thus placing most representative, most diverse and most commonly used by the user items highest in the corresponding ranking. Finally, we aggregate the $3$ rankings and select the single highest scoring item to be added to the user summary. To this end, we deploy the Borda method for aggregation of the three rankings \cite{de1781memoireunders}.

\begin{algorithm}[t!]
\begin{algorithmic}[1]
\Procedure{Profile}{$U, nn$}	
	\Comment{Profile of $U$ with at most $nn$ items }
\State\label{alg-summary-step-possible} 
$P \gets W_U \cup C_U \cup R_U$	

\State $S \gets \emptyset $

\While{$ |S| < nn $ AND $ P \not= \emptyset$ }	  \For{ $i \in P$ }
    \State $SU_i = U_i $
    	\Comment Number of times it was used by $U$ 
    \State $SR_i = \sum_{ j \in P} dist(i, j)$
    	\Comment Representative for the set 
    \State $SD_i = \sum_{j \in S} dist(i, j)$
    	\Comment Diverse based on the set 
  \EndFor

\State $r_1 \gets argsort(\mathbf{SU}, descending) $
\State $r_2 \gets argsort(\mathbf{SR}, ascending) $
\State $r_3 \gets argsort(\mathbf{SD}, descending) $

\item[]
\State $r_{final} \gets AggregateRankings(r_1, r_2, r_3) $
\State $ elem_{top} \gets Top1(r_{final})$
\State $ P \gets P \setminus \{elem_{top}\} $
\State $ S \gets S \cup \{elem_{top}\} $

\EndWhile\label{euclidendwhile}
\State \textbf{return} $S$
	\Comment{The profile consists of the items in $S$ }
\EndProcedure
\end{algorithmic}
\caption{Generating User Profiles}\label{alg:profiles}
\end{algorithm}

Extending this summarization approach to generating community rather than user profiles is a straightforward task. The main difference between the two of them is in Step \ref{alg-summary-step-possible} of Algorithm~\ref{alg:profiles} -- rather than collecting all items related to a single user, we simply aggregate all items used by any user within the community.
Finally, it is important to note that our summarization approach can be extended to take into account any type of items provided that they have comparable embeddings.

\section{Evaluation Framework} 
The usefulness of the system has been evaluated using a protocol inspired by Analytic Quality \cite{zahalka2015analytic} -- a framework which has been adapted for the evaluation of a number of analytics systems in the past years  \cite{zahalka2015interactive, zahalka2018blackthorn}. The main idea behind this framework is to automatically simulate user interactions with the system in order to evaluates its performance without the bias of human judgments.

The experiments start by creating multiple artificial actors, each one of them assigned a different task inspired by the intended use of the system. For the purpose of evaluating \Name, we assign the different actors a specific subset of the users in the corresponding dataset -- a subset which a real-life user could be potentially interested to identify by using the system. Then, the artificial actor's task is to discover as many as possible of the users in their assigned subset by interacting with the system.

According to the protocol, each artificial actor starts by presenting the system with a number of positive seed examples from their own subset. Next, we simulate multiple interaction rounds with the system. At every iteration, the system ranks all users based on the previously shown examples and top $N$ are presented to the actor, where $N$ is a parameter of the system controlling the number of examples highlighted in the interface. At this step, the artificial actor truthfully marks the presented users as relevant or irrelevant, following the assumption that a real-life interacting actor would do so themselves, striving to maximize their own information gain.

For all experiments we use the same setup -- linear SVM with SGD training as the interactive classifier, $N=15$ items are presented to the actor and we set representation dimensionality to 128.

In order to fairly evaluate the performance of the system, it is important that the criteria used to form the actor's target subsets do not take into account data used in the user representation (i.e. text, or concepts extracted from the text and the visual content). Thus we only rely on metadata such as forum structure or information about replies and retweets in order to create the target subsets. More details about this process are provided in the following section, together with the analysis of the corresponding experiments.

\section{Experimental Results}

\newcommand{\mySubfigure}[2] {
  \begin{subfigure}[t!]{0.45\textwidth}
    \includegraphics[width=\columnwidth,keepaspectratio]{#1}
    \caption{#2}
  \end{subfigure}
}

\newcommand{\myfigure}[3] {
  \begin{figure}[t!]
    \includegraphics[width=\columnwidth,keepaspectratio]{#1}
    \caption{#2}\label{#3}
  \end{figure}
}

\subsection{Baseline Experiments}\label{sec:analysis-baseline}

In our first set of experiments, we compare the performance of the different user representations, i.e., we conduct experiments with $40$ artificial actors corresponding to each of the Stormfront categories and $4$ artificial actors corresponding to each of the Twitter subcollections. 
In these experiments, each actor initially presents the system with the $15$ users which produced the highest number of posts within the corresponding category or subcollection.

Figure \ref{fig:baseline-SF} shows that the multimodal user representation based on TFIDF weighting outperforms multimodal embeddings on Stormfront dataset. A possible explanation lies in a clear difference between the topics discussed in various sub-forums. Most sub-forums further feature narrowly focused discussions, which better suits TFIDF representation than the embeddings optimized for capturing a broader context. Indeed, this difference is significantly smaller in case of Twitter (cf. Figure~\ref{fig:baseline-TW}) where such focus is absent.

\myfigure{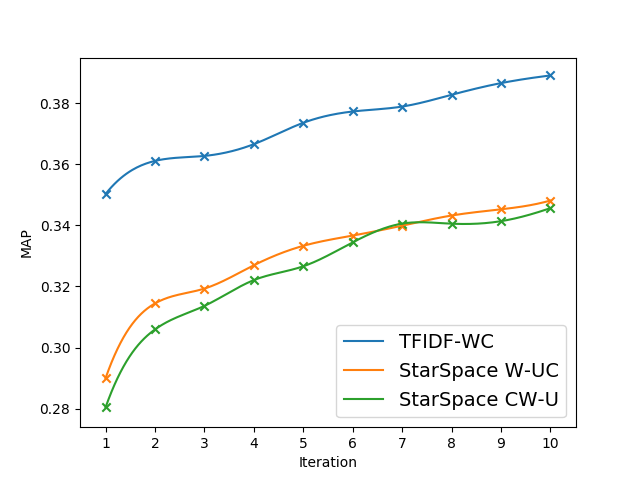}{Stormfront MAP (multimodal)}{fig:baseline-SF}
\myfigure{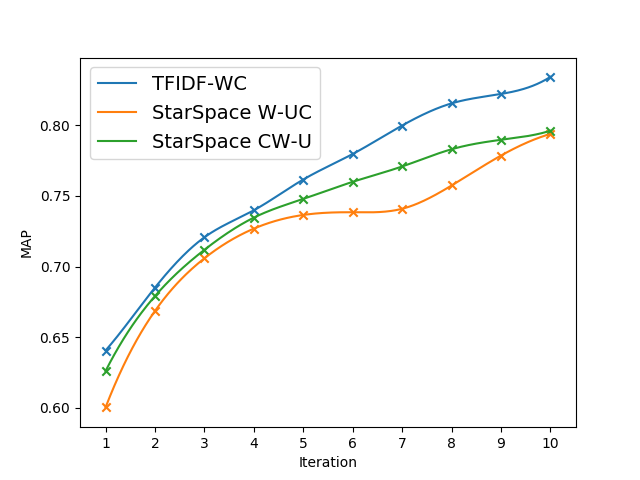}{Twitter MAP (multimodal)}{fig:baseline-TW}

\subsection{National Chapters}

Our dataset contains $14$ categories related to specific countries or geographic regions. 
These categories present a great opportunity for comparing our user representations due to the specifics of their content -- discussions cover a broad range of topics, predominantly related to history and politics, with often mentions of entities (events, people, etc) specific to the region; 
the discourse is mostly led in English, besides few chapters where the native language prevails. Thus, we next compare the performance of the different user representations in experiments with only the $14$ artificial actors corresponding to the National Chapters.

As it can be seen on Figure \ref{fig:national}, the differences in performance between the TFIDF-based representations and the multimodal embeddings is now noticeably smaller than in the experiments with all chapters. Furthermore, improvement in terms of MAP over time is considerably higher for the multimodal embeddings. It is then possible that retrieving more irrelevant results at the earlier iteration, which are truthfully labeled as such and presented as negative examples at the next round, makes it possible for the classifier based on StarSpace embeddings to learn to better distinguish between the classes.

\myfigure{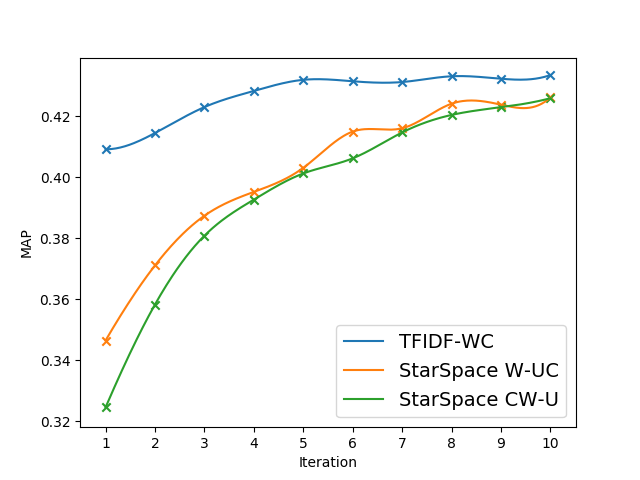}{SF National Chapters MAP (multimodal)}{fig:national}

\subsection{Individual Chapters}

Finally, the performance varies greatly even between the separate national categories, therefore, we next look into two individual categories -- \textit{Stormfront en Français}, which is characterized by the common use of the national language unlike most other national chapters, and \textit{For Stormfront Ladies Only}, which has previously been analyzed in \cite{rudinac2017multimodal} due to the interest of social scientists in the role and portrayal of women in right-wing extremist networks.

\myfigure{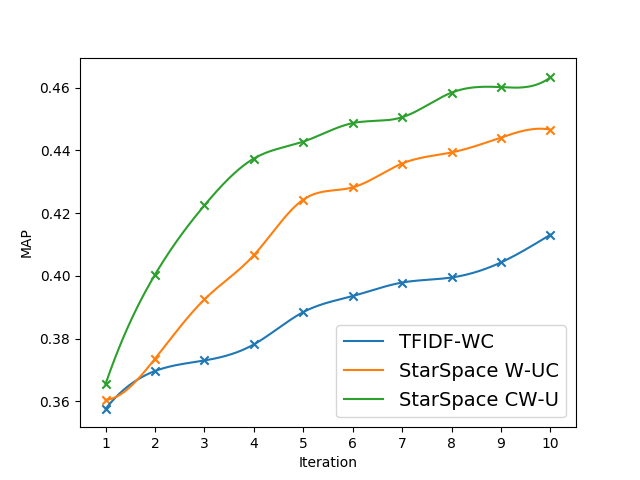}{SF Ladies MAP (multimodal)}{fig:ladies}

\myfigure{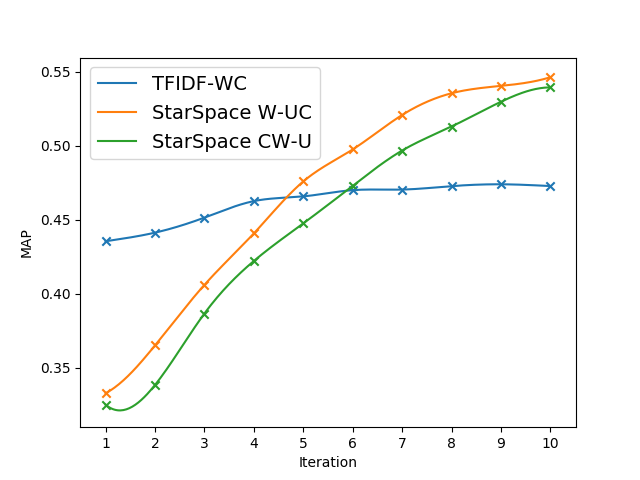}{SF France MAP (multimodal)}{fig:france}

The significantly higher results for StarSpace experiments in the experiments with the \textit{For Stormfront Ladies Only} category (cf. Figure \ref{fig:ladies}) finally prove the major advantage of the proposed approach over TFIDF representations -- that is, its ability to capture context and semantic meaning rather than encoding specific terms. \textit{For Stormfront Ladies Only} is a category which contains a vast majority of topics, including the political and ethnic discussions typical for Stormfront, but presented from the point of view of women and in relation to topics such as family, relationships and culture. This means that the specific terms and language used in the category, as well as visual concepts and entities, are overall general and similar to many other categories. This does not give TFIDF the chance `hold on' to specific terms, and highlights StarSpace's ability to capture more abstract concepts within the content.

\textit{Stormfront en Français} is one of the national chapters characterized by predominant use of the national language and a high number of extracted region-specific entities and visual concepts -- a setup which is ideal for TFIDF. However, its performance is considerably lower than that of StarSpace (cf. Figure \ref{fig:france}).

While French was hardly ever used within the same post with other languages and semantic similarities between corresponding words have not been captured by StarSpace, entities and visual concepts are more language-independent. Now they become invaluable for StarSpace, since this is the only signal which can be used to expand the topics and find more semantically similar user. Their importance can also be seen in the consistently higher scores for StarSpace W-UC, which so far performed similarly to or worse than StarSpace CW-U. 
Treating concepts as labels in the training setup (as in StarSpace W-UC) implicitly minimizes the distance between users and concepts, giving them more weight. Thus, using similar concepts becomes more important than using similar words when judging the similarity between users. This allows StarSpace W-UC to diversify the most relevant users at each iteration, helping the classifier to refine the ranking and achieve higher overall results.

\section{conclusion}

In this work, we proposed \Name{} -- a novel analytics system which facilitates gaining deeper insight into the role of a user or a group in an online community. \Name enables search and exploration in large heterogeneous collections, on the fly categorization of social multimedia users and the discovery of new users of interest,
in order to assist annotating large datasets for frequently changing communities.

As the core component of this system, we proposed a new approach to simultaneously learn multimodal representations of social multimedia users and content using the neural embedding model StarSpace.
We showed these representations to be useful not only for categorizing users, but also for automatically generating user and community profiles by selecting diverse and representative content from all modalities.

We conducted various experiments in order to automatically evaluate the performance of the system, and to identify the strengths and weaknesses of the proposed user representations.
In these experiments, we showcased StarSpace's ability to diversify the results retrieved by the system, ensuring the constant improvement of the model and the high overall long-term information gain for the user. 
We also saw that this trait of the StarSpace embeddings makes them invaluable for tasks where context is more important than the specific terms which are usually captured by TFIDF representations.

While we see \Names current implementation as a prototype of a good analytics system serving the needs of social scientists in analyzing large social multimedia networks, we have identified a number of parts of the system and its components which would benefit from future improvement.

\bibliographystyle{ACM-Reference-Format}

\bibliography{isolde}


\begin{thebibliography}{40}


\ifx \showCODEN    \undefined \def \showCODEN     #1{\unskip}     \fi
\ifx \showDOI      \undefined \def \showDOI       #1{#1}\fi
\ifx \showISBNx    \undefined \def \showISBNx     #1{\unskip}     \fi
\ifx \showISBNxiii \undefined \def \showISBNxiii  #1{\unskip}     \fi
\ifx \showISSN     \undefined \def \showISSN      #1{\unskip}     \fi
\ifx \showLCCN     \undefined \def \showLCCN      #1{\unskip}     \fi
\ifx \shownote     \undefined \def \shownote      #1{#1}          \fi
\ifx \showarticletitle \undefined \def \showarticletitle #1{#1}   \fi
\ifx \showURL      \undefined \def \showURL       {\relax}        \fi
\providecommand\bibfield[2]{#2}
\providecommand\bibinfo[2]{#2}
\providecommand\natexlab[1]{#1}
\providecommand\showeprint[2][]{arXiv:#2}

\bibitem[\protect\citeauthoryear{Barthel, Hezel, and Mackowiak}{Barthel
  et~al\mbox{.}}{2016}]%
        {10.1007/978-3-319-27674-8_43}
\bibfield{author}{\bibinfo{person}{Kai~Uwe Barthel}, \bibinfo{person}{Nico
  Hezel}, {and} \bibinfo{person}{Radek Mackowiak}.}
  \bibinfo{year}{2016}\natexlab{}.
\newblock \showarticletitle{Navigating a Graph of Scenes for Exploring Large
  Video Collections}. In \bibinfo{booktitle}{\emph{MultiMedia Modeling}},
  \bibfield{editor}{\bibinfo{person}{Qi~Tian}, \bibinfo{person}{Nicu Sebe},
  \bibinfo{person}{Guo-Jun Qi}, \bibinfo{person}{Benoit Huet},
  \bibinfo{person}{Richang Hong}, {and} \bibinfo{person}{Xueliang Liu}} (Eds.).
  \bibinfo{publisher}{Springer International Publishing},
  \bibinfo{address}{Cham}, \bibinfo{pages}{418--423}.
\newblock
\showISBNx{978-3-319-27674-8}


\bibitem[\protect\citeauthoryear{Bastian, Heymann, Jacomy,
  et~al\mbox{.}}{Bastian et~al\mbox{.}}{2009}]%
        {bastian2009gephi}
\bibfield{author}{\bibinfo{person}{Mathieu Bastian}, \bibinfo{person}{Sebastien
  Heymann}, \bibinfo{person}{Mathieu Jacomy}, {et~al\mbox{.}}}
  \bibinfo{year}{2009}\natexlab{}.
\newblock \showarticletitle{{Gephi}: an open source software for exploring and
  manipulating networks}.
\newblock \bibinfo{journal}{\emph{International AAAI Conference on Web and
  Social Media}}  \bibinfo{volume}{8} (\bibinfo{year}{2009}),
  \bibinfo{pages}{361--362}.
\newblock


\bibitem[\protect\citeauthoryear{Blei, Ng, and Jordan}{Blei
  et~al\mbox{.}}{2003}]%
        {blei2003latent}
\bibfield{author}{\bibinfo{person}{David~M Blei}, \bibinfo{person}{Andrew~Y
  Ng}, {and} \bibinfo{person}{Michael~I Jordan}.}
  \bibinfo{year}{2003}\natexlab{}.
\newblock \showarticletitle{Latent dirichlet allocation}.
\newblock \bibinfo{journal}{\emph{Journal of machine Learning research}}
  \bibinfo{volume}{3}, \bibinfo{number}{Jan} (\bibinfo{year}{2003}),
  \bibinfo{pages}{993--1022}.
\newblock


\bibitem[\protect\citeauthoryear{Bojanowski, Grave, Joulin, and
  Mikolov}{Bojanowski et~al\mbox{.}}{2017}]%
        {fasttext:2017}
\bibfield{author}{\bibinfo{person}{Piotr Bojanowski}, \bibinfo{person}{Edouard
  Grave}, \bibinfo{person}{Armand Joulin}, {and} \bibinfo{person}{Tomas
  Mikolov}.} \bibinfo{year}{2017}\natexlab{}.
\newblock \showarticletitle{Enriching Word Vectors with Subword Information}.
\newblock \bibinfo{journal}{\emph{Transactions of the Association for
  Computational Linguistics}}  \bibinfo{volume}{5} (\bibinfo{year}{2017}),
  \bibinfo{pages}{135--146}.
\newblock
\showeprint{https://doi.org/10.1162/tacl\_a\_00051}
\urldef\tempurl%
\url{https://doi.org/10.1162/tacl\_a\_00051}
\showURL{%
\tempurl}


\bibitem[\protect\citeauthoryear{Bron, Van~Gorp, Nack, Baltussen, and
  de~Rijke}{Bron et~al\mbox{.}}{2013}]%
        {comerda}
\bibfield{author}{\bibinfo{person}{Marc Bron}, \bibinfo{person}{Jasmijn
  Van~Gorp}, \bibinfo{person}{Frank Nack}, \bibinfo{person}{Lotte~Belice
  Baltussen}, {and} \bibinfo{person}{Maarten de Rijke}.}
  \bibinfo{year}{2013}\natexlab{}.
\newblock \showarticletitle{Aggregated search interface preferences in
  multi-session search tasks}. In \bibinfo{booktitle}{\emph{Proceedings of the
  36th international ACM SIGIR conference on Research and development in
  information retrieval}}. ACM, \bibinfo{pages}{123--132}.
\newblock


\bibitem[\protect\citeauthoryear{Chen, Ngo, Feng, and Chua}{Chen
  et~al\mbox{.}}{2018}]%
        {Chen:2018:DUC:3240508.3240627}
\bibfield{author}{\bibinfo{person}{Jing-Jing Chen}, \bibinfo{person}{Chong-Wah
  Ngo}, \bibinfo{person}{Fu-Li Feng}, {and} \bibinfo{person}{Tat-Seng Chua}.}
  \bibinfo{year}{2018}\natexlab{}.
\newblock \showarticletitle{Deep Understanding of Cooking Procedure for
  Cross-modal Recipe Retrieval}. In \bibinfo{booktitle}{\emph{Proceedings of
  the 26th ACM International Conference on Multimedia}}
  \emph{(\bibinfo{series}{MM '18})}. \bibinfo{publisher}{ACM},
  \bibinfo{address}{New York, NY, USA}, \bibinfo{pages}{1020--1028}.
\newblock
\showISBNx{978-1-4503-5665-7}
\urldef\tempurl%
\url{https://doi.org/10.1145/3240508.3240627}
\showDOI{\tempurl}


\bibitem[\protect\citeauthoryear{Conway}{Conway}{2016}]%
        {conway2016determining}
\bibfield{author}{\bibinfo{person}{Maura Conway}.}
  \bibinfo{year}{2016}\natexlab{}.
\newblock \showarticletitle{Determining the Role of the Internet in Violent
  Extremism and Terrorism}.
\newblock \bibinfo{journal}{\emph{Violent Extremism Online: New Perspectives on
  Terrorism and the Internet}} (\bibinfo{year}{2016}), \bibinfo{pages}{123}.
\newblock


\bibitem[\protect\citeauthoryear{de~Borda}{de~Borda}{1781}]%
        {de1781memoireunders}
\bibfield{author}{\bibinfo{person}{Jean~C de Borda}.}
  \bibinfo{year}{1781}\natexlab{}.
\newblock \showarticletitle{M{\'e}moire sur les {\'e}lections au scrutin}.
\newblock  (\bibinfo{year}{1781}).
\newblock


\bibitem[\protect\citeauthoryear{Ellson, Gansner, Koutsofios, North, and
  Woodhull}{Ellson et~al\mbox{.}}{2004}]%
        {ellson2004graphviz}
\bibfield{author}{\bibinfo{person}{John Ellson}, \bibinfo{person}{Emden~R
  Gansner}, \bibinfo{person}{Eleftherios Koutsofios},
  \bibinfo{person}{Stephen~C North}, {and} \bibinfo{person}{Gordon Woodhull}.}
  \bibinfo{year}{2004}\natexlab{}.
\newblock \showarticletitle{{Graphviz} and {Dynagraph}---Static and Dynamic
  Graph Drawing Tools}.
\newblock In \bibinfo{booktitle}{\emph{Graph drawing software}}.
  \bibinfo{publisher}{Springer}, \bibinfo{pages}{127--148}.
\newblock


\bibitem[\protect\citeauthoryear{Graus, Odijk, Tsagkias, Weerkamp, and
  De~Rijke}{Graus et~al\mbox{.}}{2014}]%
        {graus2014semanticizing}
\bibfield{author}{\bibinfo{person}{David Graus}, \bibinfo{person}{Daan Odijk},
  \bibinfo{person}{Manos Tsagkias}, \bibinfo{person}{Wouter Weerkamp}, {and}
  \bibinfo{person}{Maarten De~Rijke}.} \bibinfo{year}{2014}\natexlab{}.
\newblock \showarticletitle{Semanticizing search engine queries: the University
  of Amsterdam at the ERD 2014 challenge}. In
  \bibinfo{booktitle}{\emph{Proceedings of the first international workshop on
  Entity recognition \& disambiguation}}. ACM, \bibinfo{pages}{69--74}.
\newblock


\bibitem[\protect\citeauthoryear{Grover and Leskovec}{Grover and
  Leskovec}{2016}]%
        {Grover:2016:NSF:2939672.2939754}
\bibfield{author}{\bibinfo{person}{Aditya Grover} {and} \bibinfo{person}{Jure
  Leskovec}.} \bibinfo{year}{2016}\natexlab{}.
\newblock \showarticletitle{Node2Vec: Scalable Feature Learning for Networks}.
  In \bibinfo{booktitle}{\emph{Proceedings of the 22Nd ACM SIGKDD International
  Conference on Knowledge Discovery and Data Mining}}
  \emph{(\bibinfo{series}{KDD '16})}. \bibinfo{publisher}{ACM},
  \bibinfo{address}{New York, NY, USA}, \bibinfo{pages}{855--864}.
\newblock
\showISBNx{978-1-4503-4232-2}
\urldef\tempurl%
\url{https://doi.org/10.1145/2939672.2939754}
\showDOI{\tempurl}


\bibitem[\protect\citeauthoryear{Loko{\v{c}}, Koval{\v{c}}{\'i}k, and
  Sou{\v{c}}ek}{Loko{\v{c}} et~al\mbox{.}}{2018}]%
        {10.1007/978-3-319-73600-6_44}
\bibfield{author}{\bibinfo{person}{Jakub Loko{\v{c}}}, \bibinfo{person}{Gregor
  Koval{\v{c}}{\'i}k}, {and} \bibinfo{person}{Tom{\'a}{\v{s}} Sou{\v{c}}ek}.}
  \bibinfo{year}{2018}\natexlab{}.
\newblock \showarticletitle{Revisiting SIRET Video Retrieval Tool}. In
  \bibinfo{booktitle}{\emph{MultiMedia Modeling}},
  \bibfield{editor}{\bibinfo{person}{Klaus Schoeffmann},
  \bibinfo{person}{Thanarat~H. Chalidabhongse}, \bibinfo{person}{Chong~Wah
  Ngo}, \bibinfo{person}{Supavadee Aramvith}, \bibinfo{person}{Noel~E.
  O'Connor}, \bibinfo{person}{Yo-Sung Ho}, \bibinfo{person}{Moncef Gabbouj},
  {and} \bibinfo{person}{Ahmed Elgammal}} (Eds.). \bibinfo{publisher}{Springer
  International Publishing}, \bibinfo{address}{Cham},
  \bibinfo{pages}{419--424}.
\newblock
\showISBNx{978-3-319-73600-6}


\bibitem[\protect\citeauthoryear{{Lokoč}, {Bailer}, {Schoeffmann}, {Muenzer},
  and {Awad}}{{Lokoč} et~al\mbox{.}}{2018}]%
        {8352047}
\bibfield{author}{\bibinfo{person}{J. {Lokoč}}, \bibinfo{person}{W. {Bailer}},
  \bibinfo{person}{K. {Schoeffmann}}, \bibinfo{person}{B. {Muenzer}}, {and}
  \bibinfo{person}{G. {Awad}}.} \bibinfo{year}{2018}\natexlab{}.
\newblock \showarticletitle{On Influential Trends in Interactive Video
  Retrieval: Video Browser Showdown 2015–2017}.
\newblock \bibinfo{journal}{\emph{IEEE Transactions on Multimedia}}
  \bibinfo{volume}{20}, \bibinfo{number}{12} (\bibinfo{date}{Dec}
  \bibinfo{year}{2018}), \bibinfo{pages}{3361--3376}.
\newblock
\showISSN{1520-9210}
\urldef\tempurl%
\url{https://doi.org/10.1109/TMM.2018.2830110}
\showDOI{\tempurl}


\bibitem[\protect\citeauthoryear{Martin and Maes}{Martin and Maes}{1979}]%
        {martin1979multivariate}
\bibfield{author}{\bibinfo{person}{Nick Martin} {and} \bibinfo{person}{Hermine
  Maes}.} \bibinfo{year}{1979}\natexlab{}.
\newblock \bibinfo{booktitle}{\emph{Multivariate analysis}}.
\newblock \bibinfo{publisher}{Academic press}.
\newblock


\bibitem[\protect\citeauthoryear{Meij, Weerkamp, and de~Rijke}{Meij
  et~al\mbox{.}}{2012}]%
        {meij2012adding}
\bibfield{author}{\bibinfo{person}{Edgar Meij}, \bibinfo{person}{Wouter
  Weerkamp}, {and} \bibinfo{person}{Maarten de Rijke}.}
  \bibinfo{year}{2012}\natexlab{}.
\newblock \showarticletitle{Adding semantics to microblog posts}. In
  \bibinfo{booktitle}{\emph{Proceedings of the fifth ACM international
  conference on Web search and data mining}}. ACM, \bibinfo{pages}{563--572}.
\newblock


\bibitem[\protect\citeauthoryear{Mikolov, Sutskever, Chen, Corrado, and
  Dean}{Mikolov et~al\mbox{.}}{2013}]%
        {NIPS2013_5021}
\bibfield{author}{\bibinfo{person}{Tomas Mikolov}, \bibinfo{person}{Ilya
  Sutskever}, \bibinfo{person}{Kai Chen}, \bibinfo{person}{Greg~S Corrado},
  {and} \bibinfo{person}{Jeff Dean}.} \bibinfo{year}{2013}\natexlab{}.
\newblock \showarticletitle{Distributed Representations of Words and Phrases
  and their Compositionality}.
\newblock In \bibinfo{booktitle}{\emph{Advances in Neural Information
  Processing Systems 26}}, \bibfield{editor}{\bibinfo{person}{C.~J.~C. Burges},
  \bibinfo{person}{L.~Bottou}, \bibinfo{person}{M.~Welling},
  \bibinfo{person}{Z.~Ghahramani}, {and} \bibinfo{person}{K.~Q. Weinberger}}
  (Eds.). \bibinfo{publisher}{Curran Associates, Inc.},
  \bibinfo{pages}{3111--3119}.
\newblock
\urldef\tempurl%
\url{http://papers.nips.cc/paper/5021-distributed-representations-of-words-and-phrases-and-their-compositionality.pdf}
\showURL{%
\tempurl}


\bibitem[\protect\citeauthoryear{Odijk}{Odijk}{2012}]%
        {semanticizer:2012}
\bibfield{author}{\bibinfo{person}{Daan Odijk}.}
  \bibinfo{year}{2012}\natexlab{}.
\newblock \bibinfo{title}{{UvA Semanticizer Web API}}.
\newblock
  \bibinfo{howpublished}{\url{https://github.com/semanticize/semanticizer}}.
\newblock


\bibitem[\protect\citeauthoryear{Pennington, Socher, and Manning}{Pennington
  et~al\mbox{.}}{2014}]%
        {D14-1162}
\bibfield{author}{\bibinfo{person}{Jeffrey Pennington},
  \bibinfo{person}{Richard Socher}, {and} \bibinfo{person}{Christopher
  Manning}.} \bibinfo{year}{2014}\natexlab{}.
\newblock \showarticletitle{Glove: Global Vectors for Word Representation}. In
  \bibinfo{booktitle}{\emph{Proceedings of the 2014 Conference on Empirical
  Methods in Natural Language Processing (EMNLP)}}.
  \bibinfo{publisher}{Association for Computational Linguistics},
  \bibinfo{address}{Doha, Qatar}, \bibinfo{pages}{1532--1543}.
\newblock
\urldef\tempurl%
\url{https://doi.org/10.3115/v1/D14-1162}
\showDOI{\tempurl}


\bibitem[\protect\citeauthoryear{Qi, Wang, and Li}{Qi et~al\mbox{.}}{2017}]%
        {Qi:2017:OCS:3123266.3123311}
\bibfield{author}{\bibinfo{person}{Mengshi Qi}, \bibinfo{person}{Yunhong Wang},
  {and} \bibinfo{person}{Annan Li}.} \bibinfo{year}{2017}\natexlab{}.
\newblock \showarticletitle{Online Cross-Modal Scene Retrieval by Binary
  Representation and Semantic Graph}. In \bibinfo{booktitle}{\emph{Proceedings
  of the 25th ACM International Conference on Multimedia}}
  \emph{(\bibinfo{series}{MM '17})}. \bibinfo{publisher}{ACM},
  \bibinfo{address}{New York, NY, USA}, \bibinfo{pages}{744--752}.
\newblock
\showISBNx{978-1-4503-4906-2}
\urldef\tempurl%
\url{https://doi.org/10.1145/3123266.3123311}
\showDOI{\tempurl}


\bibitem[\protect\citeauthoryear{Rossetto, Amiri~Parian, Gasser, Giangreco,
  Heller, and Schuldt}{Rossetto et~al\mbox{.}}{2019}]%
        {10.1007/978-3-030-05716-9_55}
\bibfield{author}{\bibinfo{person}{Luca Rossetto}, \bibinfo{person}{Mahnaz
  Amiri~Parian}, \bibinfo{person}{Ralph Gasser}, \bibinfo{person}{Ivan
  Giangreco}, \bibinfo{person}{Silvan Heller}, {and} \bibinfo{person}{Heiko
  Schuldt}.} \bibinfo{year}{2019}\natexlab{}.
\newblock \showarticletitle{Deep Learning-Based Concept Detection in vitrivr}.
  In \bibinfo{booktitle}{\emph{MultiMedia Modeling}},
  \bibfield{editor}{\bibinfo{person}{Ioannis Kompatsiaris},
  \bibinfo{person}{Benoit Huet}, \bibinfo{person}{Vasileios Mezaris},
  \bibinfo{person}{Cathal Gurrin}, \bibinfo{person}{Wen-Huang Cheng}, {and}
  \bibinfo{person}{Stefanos Vrochidis}} (Eds.). \bibinfo{publisher}{Springer
  International Publishing}, \bibinfo{address}{Cham},
  \bibinfo{pages}{616--621}.
\newblock
\showISBNx{978-3-030-05716-9}


\bibitem[\protect\citeauthoryear{Rudinac, Chua, Diaz-Ferreyra, Friedland,
  Gornostaja, Huet, Kaptein, Lind{\'e}n, Moens, Peltonen,
  et~al\mbox{.}}{Rudinac et~al\mbox{.}}{2018}]%
        {rudinac2018rethinking}
\bibfield{author}{\bibinfo{person}{Stevan Rudinac}, \bibinfo{person}{Tat-Seng
  Chua}, \bibinfo{person}{Nicolas Diaz-Ferreyra}, \bibinfo{person}{Gerald
  Friedland}, \bibinfo{person}{Tatjana Gornostaja}, \bibinfo{person}{Benoit
  Huet}, \bibinfo{person}{Rianne Kaptein}, \bibinfo{person}{Krister
  Lind{\'e}n}, \bibinfo{person}{Marie-Francine Moens}, \bibinfo{person}{Jaakko
  Peltonen}, {et~al\mbox{.}}} \bibinfo{year}{2018}\natexlab{}.
\newblock \showarticletitle{Rethinking Summarization and Storytelling for
  Modern Social Multimedia}.
\newblock \bibinfo{journal}{\emph{Lecture Notes in Computer Science}}
  \bibinfo{volume}{10704} (\bibinfo{year}{2018}).
\newblock


\bibitem[\protect\citeauthoryear{Rudinac, Gornishka, and Worring}{Rudinac
  et~al\mbox{.}}{2017}]%
        {rudinac2017multimodal}
\bibfield{author}{\bibinfo{person}{Stevan Rudinac}, \bibinfo{person}{Iva
  Gornishka}, {and} \bibinfo{person}{Marcel Worring}.}
  \bibinfo{year}{2017}\natexlab{}.
\newblock \showarticletitle{Multimodal Classification of Violent Online
  Political Extremism Content with Graph Convolutional Networks}. In
  \bibinfo{booktitle}{\emph{Proceedings of the on Thematic Workshops of ACM
  Multimedia 2017}}. ACM, \bibinfo{pages}{245--252}.
\newblock


\bibitem[\protect\citeauthoryear{Salton and Buckley}{Salton and
  Buckley}{1988}]%
        {salton1988term}
\bibfield{author}{\bibinfo{person}{Gerard Salton} {and}
  \bibinfo{person}{Christopher Buckley}.} \bibinfo{year}{1988}\natexlab{}.
\newblock \showarticletitle{Term-weighting approaches in automatic text
  retrieval}.
\newblock \bibinfo{journal}{\emph{Information processing \& management}}
  \bibinfo{volume}{24}, \bibinfo{number}{5} (\bibinfo{year}{1988}),
  \bibinfo{pages}{513--523}.
\newblock


\bibitem[\protect\citeauthoryear{Smith, Shneiderman, Milic-Frayling,
  Mendes~Rodrigues, Barash, Dunne, Capone, Perer, and Gleave}{Smith
  et~al\mbox{.}}{2009}]%
        {smith2009analyzing}
\bibfield{author}{\bibinfo{person}{Marc~A Smith}, \bibinfo{person}{Ben
  Shneiderman}, \bibinfo{person}{Natasa Milic-Frayling},
  \bibinfo{person}{Eduarda Mendes~Rodrigues}, \bibinfo{person}{Vladimir
  Barash}, \bibinfo{person}{Cody Dunne}, \bibinfo{person}{Tony Capone},
  \bibinfo{person}{Adam Perer}, {and} \bibinfo{person}{Eric Gleave}.}
  \bibinfo{year}{2009}\natexlab{}.
\newblock \showarticletitle{Analyzing (social media) networks with {NodeXL}}.
  In \bibinfo{booktitle}{\emph{Proceedings of the fourth International
  Conference on Communities and Technologies}}. ACM, \bibinfo{pages}{255--264}.
\newblock


\bibitem[\protect\citeauthoryear{Snoek, van~de Sande, Fontijne, Habibian, Jain,
  Kordumova, Li, Mazloom, Pintea, Tao, Koelma, and Smeulders}{Snoek
  et~al\mbox{.}}{2013}]%
        {SnoekPTRECVID2013}
\bibfield{author}{\bibinfo{person}{C.~G.~M. Snoek}, \bibinfo{person}{K.~E.~A.
  van~de Sande}, \bibinfo{person}{D. Fontijne}, \bibinfo{person}{A. Habibian},
  \bibinfo{person}{M. Jain}, \bibinfo{person}{S. Kordumova},
  \bibinfo{person}{Z. Li}, \bibinfo{person}{M. Mazloom}, \bibinfo{person}{S.~L.
  Pintea}, \bibinfo{person}{R. Tao}, \bibinfo{person}{D.~C. Koelma}, {and}
  \bibinfo{person}{A.~W.~M. Smeulders}.} \bibinfo{year}{2013}\natexlab{}.
\newblock \showarticletitle{MediaMill at TRECVID 2013: Searching Concepts,
  Objects, Instances and Events in Video}. In \bibinfo{booktitle}{\emph{TRECVID
  Workshop}}.
\newblock
\urldef\tempurl%
\url{https://ivi.fnwi.uva.nl/isis/publications/2013/SnoekPTRECVID2013}
\showURL{%
\tempurl}


\bibitem[\protect\citeauthoryear{Strohman, Metzler, Turtle, and Croft}{Strohman
  et~al\mbox{.}}{2005}]%
        {strohman2005indri}
\bibfield{author}{\bibinfo{person}{Trevor Strohman}, \bibinfo{person}{Donald
  Metzler}, \bibinfo{person}{Howard Turtle}, {and} \bibinfo{person}{W~Bruce
  Croft}.} \bibinfo{year}{2005}\natexlab{}.
\newblock \showarticletitle{{Indri}: A language model-based search engine for
  complex queries}. In \bibinfo{booktitle}{\emph{Proceedings of the
  International Conference on Intelligent Analysis}}, Vol.~\bibinfo{volume}{2}.
  Citeseer, \bibinfo{pages}{2--6}.
\newblock


\bibitem[\protect\citeauthoryear{Teevan, Ramage, and Morris}{Teevan
  et~al\mbox{.}}{2011}]%
        {teevan2011twittersearch}
\bibfield{author}{\bibinfo{person}{Jaime Teevan}, \bibinfo{person}{Daniel
  Ramage}, {and} \bibinfo{person}{Merredith~Ringel Morris}.}
  \bibinfo{year}{2011}\natexlab{}.
\newblock \showarticletitle{\# TwitterSearch: a comparison of microblog search
  and web search}. In \bibinfo{booktitle}{\emph{Proceedings of the fourth ACM
  international conference on Web search and data mining}}. ACM,
  \bibinfo{pages}{35--44}.
\newblock


\bibitem[\protect\citeauthoryear{van~der Corput and van Wijk}{van~der Corput
  and van Wijk}{2016}]%
        {van2016iclic}
\bibfield{author}{\bibinfo{person}{Paul van~der Corput} {and}
  \bibinfo{person}{Jarke~J van Wijk}.} \bibinfo{year}{2016}\natexlab{}.
\newblock \showarticletitle{{ICLIC}: Interactive categorization of large image
  collections}. In \bibinfo{booktitle}{\emph{Pacific Visualization Symposium
  (PacificVis), 2016 IEEE}}. IEEE, \bibinfo{pages}{152--159}.
\newblock


\bibitem[\protect\citeauthoryear{van~der Corput and van Wijk}{van~der Corput
  and van Wijk}{2017}]%
        {doi:10.1111/cgf.13188}
\bibfield{author}{\bibinfo{person}{Paul van~der Corput} {and}
  \bibinfo{person}{Jarke~J. van Wijk}.} \bibinfo{year}{2017}\natexlab{}.
\newblock \showarticletitle{Comparing Personal Image Collections with
  PICTuReVis}.
\newblock \bibinfo{journal}{\emph{Computer Graphics Forum}}
  \bibinfo{volume}{36}, \bibinfo{number}{3} (\bibinfo{year}{2017}),
  \bibinfo{pages}{295--304}.
\newblock
\urldef\tempurl%
\url{https://doi.org/10.1111/cgf.13188}
\showDOI{\tempurl}
\showeprint{https://onlinelibrary.wiley.com/doi/pdf/10.1111/cgf.13188}


\bibitem[\protect\citeauthoryear{Wang, Yang, Xu, Hanjalic, and Shen}{Wang
  et~al\mbox{.}}{2017}]%
        {Wang:2017:ACR:3123266.3123326}
\bibfield{author}{\bibinfo{person}{Bokun Wang}, \bibinfo{person}{Yang Yang},
  \bibinfo{person}{Xing Xu}, \bibinfo{person}{Alan Hanjalic}, {and}
  \bibinfo{person}{Heng~Tao Shen}.} \bibinfo{year}{2017}\natexlab{}.
\newblock \showarticletitle{Adversarial Cross-Modal Retrieval}. In
  \bibinfo{booktitle}{\emph{Proceedings of the 25th ACM International
  Conference on Multimedia}} \emph{(\bibinfo{series}{MM '17})}.
  \bibinfo{publisher}{ACM}, \bibinfo{address}{New York, NY, USA},
  \bibinfo{pages}{154--162}.
\newblock
\showISBNx{978-1-4503-4906-2}
\urldef\tempurl%
\url{https://doi.org/10.1145/3123266.3123326}
\showDOI{\tempurl}


\bibitem[\protect\citeauthoryear{Worring, Koelma, and Zahálka}{Worring
  et~al\mbox{.}}{2016}]%
        {7579240}
\bibfield{author}{\bibinfo{person}{Marcel Worring}, \bibinfo{person}{Dennis
  Koelma}, {and} \bibinfo{person}{Jan Zahálka}.}
  \bibinfo{year}{2016}\natexlab{}.
\newblock \showarticletitle{Multimedia Pivot Tables for Multimedia Analytics on
  Image Collections}.
\newblock \bibinfo{journal}{\emph{IEEE Transactions on Multimedia}}
  \bibinfo{volume}{18}, \bibinfo{number}{11} (\bibinfo{date}{Nov}
  \bibinfo{year}{2016}), \bibinfo{pages}{2217--2227}.
\newblock
\showISSN{1520-9210}
\urldef\tempurl%
\url{https://doi.org/10.1109/TMM.2016.2614380}
\showDOI{\tempurl}


\bibitem[\protect\citeauthoryear{Wu, Fisch, Chopra, Adams, Bordes, and
  Weston}{Wu et~al\mbox{.}}{2017}]%
        {wu2017starspace}
\bibfield{author}{\bibinfo{person}{Ledell Wu}, \bibinfo{person}{Adam Fisch},
  \bibinfo{person}{Sumit Chopra}, \bibinfo{person}{Keith Adams},
  \bibinfo{person}{Antoine Bordes}, {and} \bibinfo{person}{Jason Weston}.}
  \bibinfo{year}{2017}\natexlab{}.
\newblock \showarticletitle{{StarSpace}: Embed All The Things!}
\newblock \bibinfo{journal}{\emph{arXiv preprint arXiv:1709.03856}}
  (\bibinfo{year}{2017}).
\newblock


\bibitem[\protect\citeauthoryear{Wu, Fisch, Chopra, Adams, Bordes, and
  Weston}{Wu et~al\mbox{.}}{2018a}]%
        {DBLP:conf/aaai/WuFCABW18}
\bibfield{author}{\bibinfo{person}{Ledell~Yu Wu}, \bibinfo{person}{Adam Fisch},
  \bibinfo{person}{Sumit Chopra}, \bibinfo{person}{Keith Adams},
  \bibinfo{person}{Antoine Bordes}, {and} \bibinfo{person}{Jason Weston}.}
  \bibinfo{year}{2018}\natexlab{a}.
\newblock \showarticletitle{StarSpace: Embed All The Things!}, See
  \citeN{DBLP:conf/aaai/WuFCABW18}.
\newblock
\urldef\tempurl%
\url{https://www.aaai.org/ocs/index.php/AAAI/AAAI18/paper/view/16998}
\showURL{%
\tempurl}


\bibitem[\protect\citeauthoryear{Wu, Wang, and Huang}{Wu
  et~al\mbox{.}}{2018b}]%
        {Wu:2018:LSS:3240508.3240521}
\bibfield{author}{\bibinfo{person}{Yiling Wu}, \bibinfo{person}{Shuhui Wang},
  {and} \bibinfo{person}{Qingming Huang}.} \bibinfo{year}{2018}\natexlab{b}.
\newblock \showarticletitle{Learning Semantic Structure-preserved Embeddings
  for Cross-modal Retrieval}. In \bibinfo{booktitle}{\emph{Proceedings of the
  26th ACM International Conference on Multimedia}} \emph{(\bibinfo{series}{MM
  '18})}. \bibinfo{publisher}{ACM}, \bibinfo{address}{New York, NY, USA},
  \bibinfo{pages}{825--833}.
\newblock
\showISBNx{978-1-4503-5665-7}
\urldef\tempurl%
\url{https://doi.org/10.1145/3240508.3240521}
\showDOI{\tempurl}


\bibitem[\protect\citeauthoryear{Yang, Liu, Zhang, Yuan, Zhao, Barlowe, and
  Liu}{Yang et~al\mbox{.}}{2013}]%
        {yang2013piwi}
\bibfield{author}{\bibinfo{person}{Jing Yang}, \bibinfo{person}{Yujie Liu},
  \bibinfo{person}{Xin Zhang}, \bibinfo{person}{Xiaoru Yuan},
  \bibinfo{person}{Ye Zhao}, \bibinfo{person}{Scott Barlowe}, {and}
  \bibinfo{person}{Shixia Liu}.} \bibinfo{year}{2013}\natexlab{}.
\newblock \showarticletitle{{PIWI}: Visually exploring graphs based on their
  community structure}.
\newblock \bibinfo{journal}{\emph{IEEE Transactions on Visualization and
  Computer Graphics}} \bibinfo{volume}{19}, \bibinfo{number}{6}
  (\bibinfo{year}{2013}), \bibinfo{pages}{1034--1047}.
\newblock


\bibitem[\protect\citeauthoryear{Yang, Luo, and Liu}{Yang
  et~al\mbox{.}}{2010}]%
        {yang2010newdle}
\bibfield{author}{\bibinfo{person}{Jing Yang}, \bibinfo{person}{Dongning Luo},
  {and} \bibinfo{person}{Yujie Liu}.} \bibinfo{year}{2010}\natexlab{}.
\newblock \showarticletitle{{Newdle}: Interactive visual exploration of large
  online news collections}.
\newblock \bibinfo{journal}{\emph{IEEE computer graphics and applications}}
  \bibinfo{volume}{30}, \bibinfo{number}{5} (\bibinfo{year}{2010}),
  \bibinfo{pages}{32--41}.
\newblock


\bibitem[\protect\citeauthoryear{Yang, Luo, Chen, Shen, Shao, and Shen}{Yang
  et~al\mbox{.}}{2016}]%
        {Yang:2016:ZHV:2964284.2964319}
\bibfield{author}{\bibinfo{person}{Yang Yang}, \bibinfo{person}{Yadan Luo},
  \bibinfo{person}{Weilun Chen}, \bibinfo{person}{Fumin Shen},
  \bibinfo{person}{Jie Shao}, {and} \bibinfo{person}{Heng~Tao Shen}.}
  \bibinfo{year}{2016}\natexlab{}.
\newblock \showarticletitle{Zero-Shot Hashing via Transferring Supervised
  Knowledge}. In \bibinfo{booktitle}{\emph{Proceedings of the 24th ACM
  International Conference on Multimedia}} \emph{(\bibinfo{series}{MM '16})}.
  \bibinfo{publisher}{ACM}, \bibinfo{address}{New York, NY, USA},
  \bibinfo{pages}{1286--1295}.
\newblock
\showISBNx{978-1-4503-3603-1}
\urldef\tempurl%
\url{https://doi.org/10.1145/2964284.2964319}
\showDOI{\tempurl}


\bibitem[\protect\citeauthoryear{Zah{\'a}lka, Rudinac, J{\'o}nsson, Koelma, and
  Worring}{Zah{\'a}lka et~al\mbox{.}}{2018}]%
        {zahalka2018blackthorn}
\bibfield{author}{\bibinfo{person}{Jan Zah{\'a}lka}, \bibinfo{person}{Stevan
  Rudinac}, \bibinfo{person}{Bj{\"o}rn~Th{\'o}r J{\'o}nsson},
  \bibinfo{person}{Dennis~C Koelma}, {and} \bibinfo{person}{Marcel Worring}.}
  \bibinfo{year}{2018}\natexlab{}.
\newblock \showarticletitle{Blackthorn: Large-Scale Interactive Multimodal
  Learning}.
\newblock \bibinfo{journal}{\emph{IEEE Transactions on Multimedia}}
  \bibinfo{volume}{20}, \bibinfo{number}{3} (\bibinfo{year}{2018}),
  \bibinfo{pages}{687--698}.
\newblock


\bibitem[\protect\citeauthoryear{Zah{\'a}lka, Rudinac, and Worring}{Zah{\'a}lka
  et~al\mbox{.}}{2015a}]%
        {zahalka2015analytic}
\bibfield{author}{\bibinfo{person}{Jan Zah{\'a}lka}, \bibinfo{person}{Stevan
  Rudinac}, {and} \bibinfo{person}{Marcel Worring}.}
  \bibinfo{year}{2015}\natexlab{a}.
\newblock \showarticletitle{Analytic quality: evaluation of performance and
  insight in multimedia collection analysis}. In
  \bibinfo{booktitle}{\emph{Proceedings of the 23rd ACM international
  conference on Multimedia}}. ACM, \bibinfo{pages}{231--240}.
\newblock


\bibitem[\protect\citeauthoryear{Zah{\'a}lka, Rudinac, and Worring}{Zah{\'a}lka
  et~al\mbox{.}}{2015b}]%
        {zahalka2015interactive}
\bibfield{author}{\bibinfo{person}{Jan Zah{\'a}lka}, \bibinfo{person}{Stevan
  Rudinac}, {and} \bibinfo{person}{Marcel Worring}.}
  \bibinfo{year}{2015}\natexlab{b}.
\newblock \showarticletitle{Interactive multimodal learning for venue
  recommendation}.
\newblock \bibinfo{journal}{\emph{IEEE Transactions on Multimedia}}
  \bibinfo{volume}{17}, \bibinfo{number}{12} (\bibinfo{year}{2015}),
  \bibinfo{pages}{2235--2244}.
\newblock


\end{thebibliography}

\end{document}